\documentclass{icrc2009}

\usepackage[pdftex]{graphicx}
\usepackage [pagewise, mathlines, displaymath]{lineno}
\usepackage{subfig} % subfig.sty for a double column floating figure using two subfigures
\usepackage{fixltx2e}
\usepackage{url}

\newcommand{\shorttitle}[1]%
{\markboth{Proceedings of the 31\MakeLowercase{$^{st}$} ICRC, {\L}\'{o}d\'{z} 2009}{#1} }
\newcommand{\etal}{\MakeLowercase{\textit{et al. }}} % "et al."

\begin{document}
\title{Radio detection of cosmic rays at the southern Auger Observatory}

\author{\IEEEauthorblockN{A.M. van den Berg\IEEEauthorrefmark{1},
			  for the Pierre Auger Collaboration\IEEEauthorrefmark{2}}
                            \\
\IEEEauthorblockA{\IEEEauthorrefmark{1}Kernfysisch Versneller Instituut, University of Groningen, NL-9747 AA, Groningen, The Netherlands}
\IEEEauthorblockA{\IEEEauthorrefmark{2}Av. San Mart\'{\i}n Norte 304, (5613) Malarg\"ue, Prov. de Mendoza, Argentina}}

% please write the preseter's name and short title (3-4 words maximum)
%    which will appear at the header of the even pages.
\shorttitle{A.M. van den Berg \etal Radio detection of cosmic rays at Auger}
\maketitle

\begin{abstract}
An integrated approach has been developed to study radio signals induced by cosmic rays entering the Earth's atmosphere. An engineering array will be co-located with the infill array of the Pierre Auger Observatory. Our R\&D effort includes the physics processes leading to the development of radio signals, end-to-end simulations of realistic hardware configurations, and tests of various systems on site, where coincidences with the other detector systems of the Observatory are used to benchmark the systems under development.	
\end{abstract}

\begin{IEEEkeywords}
 radio detection
\end{IEEEkeywords}

\section{Introduction}
 
Results from the Southern Pierre Auger Observatory as well as the baseline design of the Northern Observatory \cite{AugerNorth} point to the need of very large aperture detection systems for ultra-high energy cosmic rays (UHE CRs); see, e.g., Ref \cite{Olinto2009a}. There are a number of worldwide efforts to develop and establish new detection techniques that promise a cost-effective extension of currently available apertures to even larger dimensions. These are, for example, the observation from space of fluorescence emission by showers or the use of large arrays of radio antennas.\\ 

\noindent
The detection of radio emission induced by high-energy and ultra-high-energy cosmic rays hitting the Earth's atmosphere is possible because of coherent radiation from the extensive air shower at radio frequencies. This radiation, which is emitted by secondary particles created in the air shower, can be measured with simple radio antennas, as was demonstrated first by Jelley in 1965 \cite{Jelley1965a}. Recently, improved technology has led to a revival of this technique. Radio detectors, like LOPES  and CODALEMA produce promising results at energies beyond 10$^{17}$ eV \cite{Haungs2009a, Lautrido2009a}.\\

\noindent
With its nearly 100\% duty cycle, a signal-to-noise ratio scaling with the square of the cosmic-ray energy, its high angular resolution, and its sensitivity to the longitudinal air-shower evolution, the radio technique is particularly well-suited for detection of UHE CRs in large-scale arrays. Therefore, we are performing an R\&D project to study UHE CRs using the detection of coherent radio emission from air showers in the Earth's atmosphere. The project, called AERA (Auger Engineering Radio Array), will have a dimension of about 20 km$^2$. For such an area we expect an UHE CR rate of about 5000 identified radio events per year. This data set will be used to address both scientific and technological questions. At the same time, the scale of such an array is large enough to test concepts for the deployment of hardware, for the operation of hardware and software, and to monitor the sustainability of critical parts of the whole system for a much larger array. \\

	\begin{figure}[h!t]
	\centering
	\includegraphics[width=.48\textwidth, viewport = 125 10 675 460, clip]{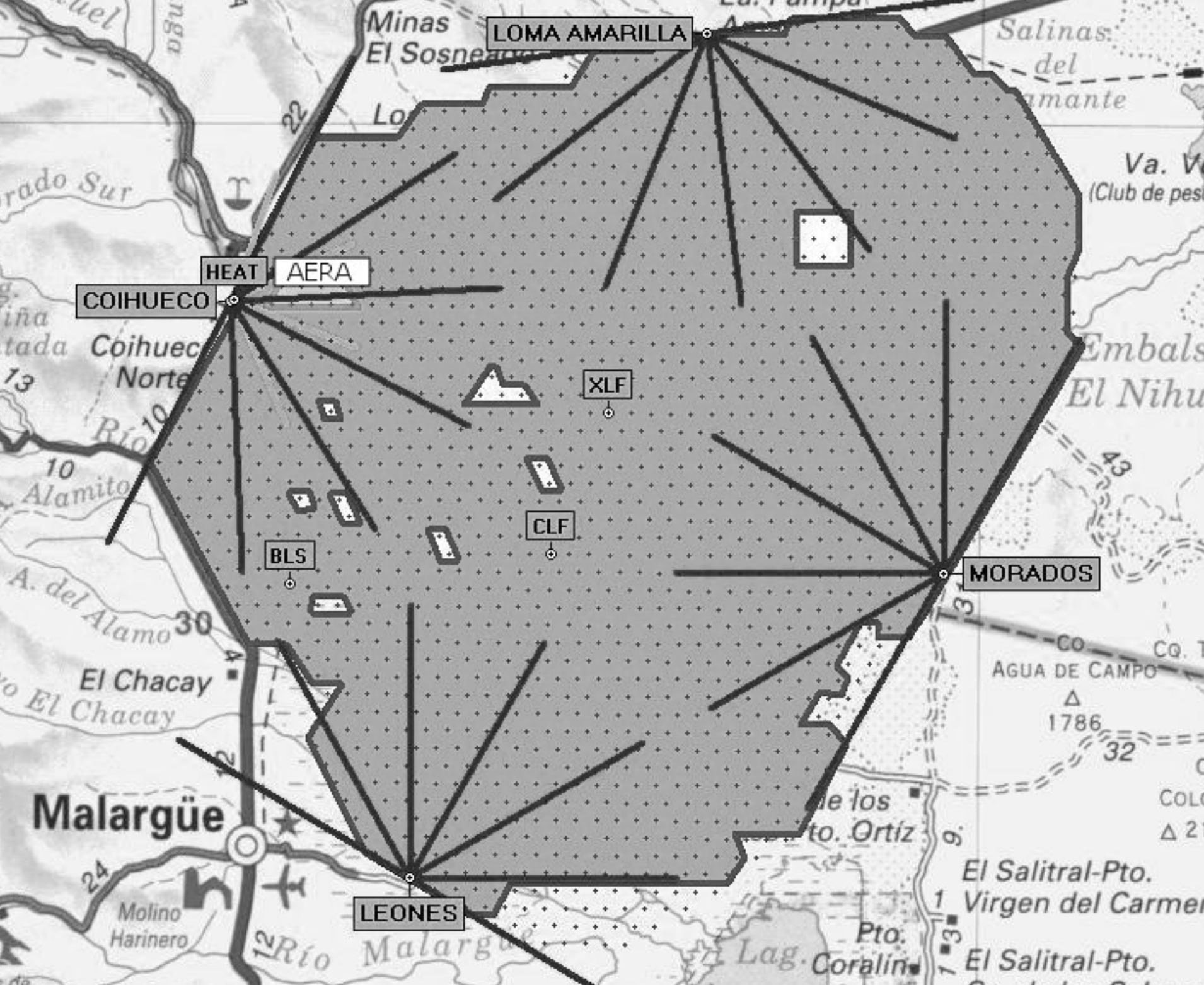}
	\caption{Locations of the radio setups at the southern Observatory. Existing setups are located near the BLS and the CLF; the future AERA is near Coihueco.}
	\label{layoutAS}
	\end{figure}

\noindent
AERA will fulfill three science goals which are not independent from each other, but have to be tackled sequentially in the time of operation and analysis and in close conjunction with data from the baseline detectors of the Observatory \cite{baseline} and the enhancements AMIGA\cite{AMIGA} and HEAT \cite{HEAT}. They are all located in the north-eastern part of the Observatory near the Coihueco fluorescence building, see Fig. \ref{layoutAS}. The three science goals are listed below.

\begin{enumerate}
\item Thorough investigation of the radio emission from an air shower at the highest energies. This includes the understanding of all the dependencies of the radio signal on general shower parameters and on the shower geometry. By this a better insight into the underlying emission mechanism will be possible; the question has to be answered which is the theory of choice and to which level one has to consider additional effects \cite{Huege2009a}. 

\item Exploration of the capability of the radio-detection technique. Determination of the extend and accuracy of stand-alone radio-detection to provide information on the most important physics quantities of UHE CRs: primary energy, primary mass, and arrival direction.

\item Composition measurements between $10^{17.4}$ and $10^{18.7}$ eV where we expect the transition from galactic to extragalactic origin of cosmic rays. With super-hybrid measurements at the AMIGA site \cite{AMIGA}, AERA will contribute with a worldwide unrivaled precision to the study of the cosmic-ray energy spectrum and composition in the energy range of the transition .
\end{enumerate}
%\parskip	

\noindent
AERA will be co-located with AMIGA, which is overlooked by the HEAT detector. These enhancements of the Observatory provide worldwide the only possibility to study the details of radio emission from air showers in a timely manner. 

\section{Present status of the project}

Initial measurements have been made showing that indeed radio detection of UHE CRs can be performed at the southern site of the Auger Observatory with a setup near the BLS, another one near the CLF (see Fig. \ref{layoutAS}). Still the number of radio events in coincidence with its Surface Detector Array (SD) is relatively small, mainly because of the relatively small scale of the radio detector arrays used. A description of the setups used is presented in Ref. \cite{berg2008}. Here we list some of the initial results and present a short description of these setups. One setup consists of three dual active fat dipole antennas, mounted near the CLF in a triangular configuration with a baseline of 139~m. With this setup, self-triggered events are being recorded and using GPS time stamps, they are compared to events registered with the SD Array of the Observatory. The other setup, located near the BLS, uses dual logarithmic periodic dipole antennas and is triggered externally using a set of two particle detectors. Also in this case the Radio-Detection Stations are mounted on a triangular grid; here the baseline is 100~m. Both setups measure the electric field strength in two polarization directions: east-west and north-south.\\

\begin{figure}[h!t]
	\centering
	\includegraphics[width=.48\textwidth]{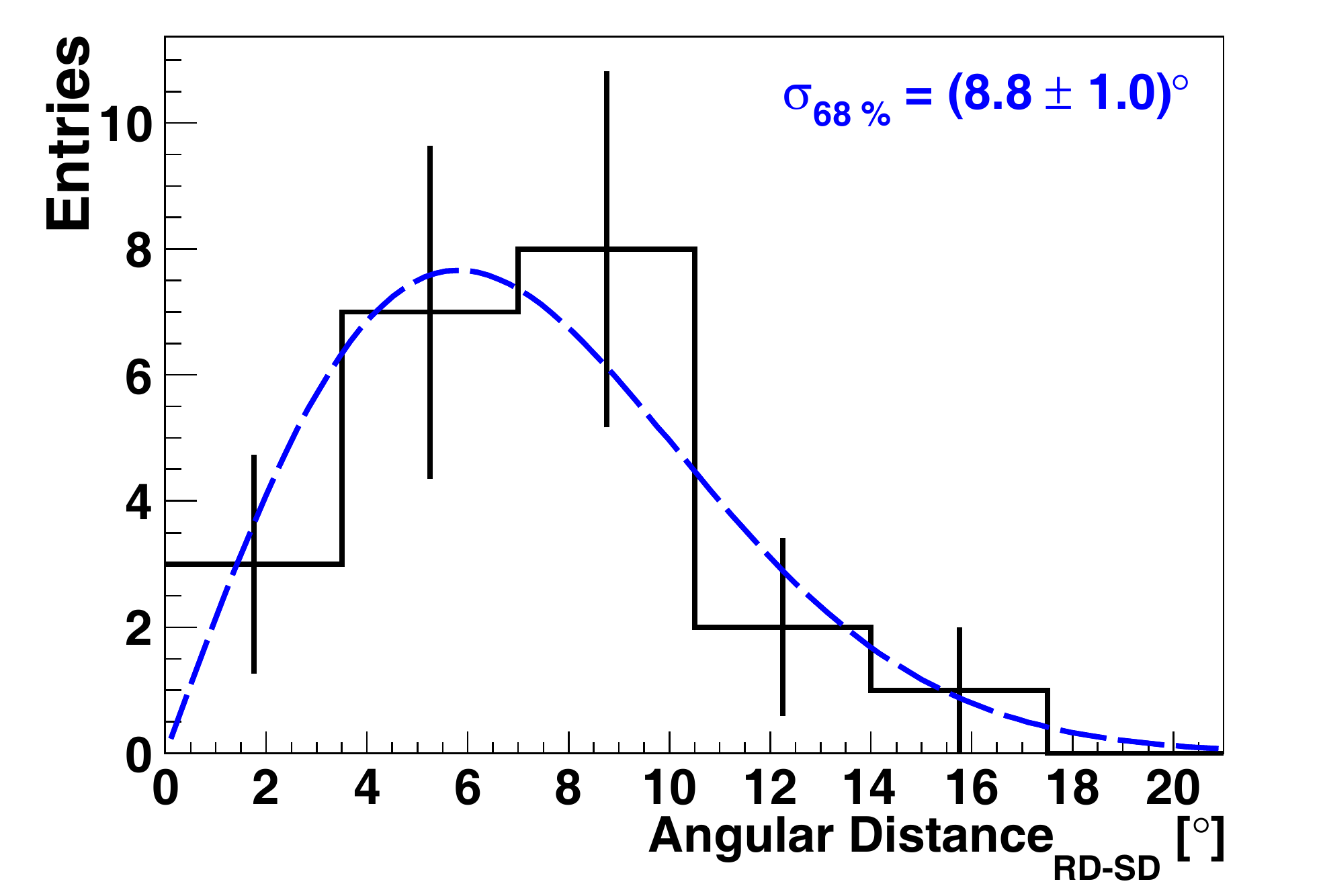}
	\caption{Plot of the angular difference between the arrival direction of observed air showers using data from the Surface Detector Array and from the Radio-Detection Stations near the BLS. The histogram indicates the data obtained for 27 events; the dashed line is a fit through these data using a Rayleigh function; see text for details.}
	\label{angles} %This pic is the opener for the Analysis part
	\end{figure}

\noindent
Several coincident events between SD and externally triggered Radio-Detection Stations located near the BLS have been recorded \cite{Coppens2008a}. In 27 cases we have recorded simultaneously in three different Radio-Detection Stations a coincidence with the SD. This allowed us to compare in these cases the arrival direction as determined by our Radio-Detection Stations with that from SD. The histogram in Fig. \ref{angles} displays for 27 events the measured angular difference between the arrival direction as determined with SD and as determined by our three Radio-Detection Stations. Using a Rayleigh function, the 68\%-quantile of the distribution of the angular difference has been determined to be $(8.8 \pm 1.0)^\circ$. In most events, the number of SD stations used in the event reconstruction was 3 or 4. The angular resolution for these type of SD events is about $2.0^\circ$ and can thus be neglected. The major contribution to the angular uncertainty is the relative small distance of 100~m between the Radio-Detection Stations compared to the timing accuracy obtained (about 3~ns).\\

%In addition, the lateral distribution function has been studied \cite{Coppens2008a} at energies higher than those at LOPES and CODALEMA.
	
\begin{figure}[h!t]
	\centering
	\includegraphics[width=.48\textwidth]{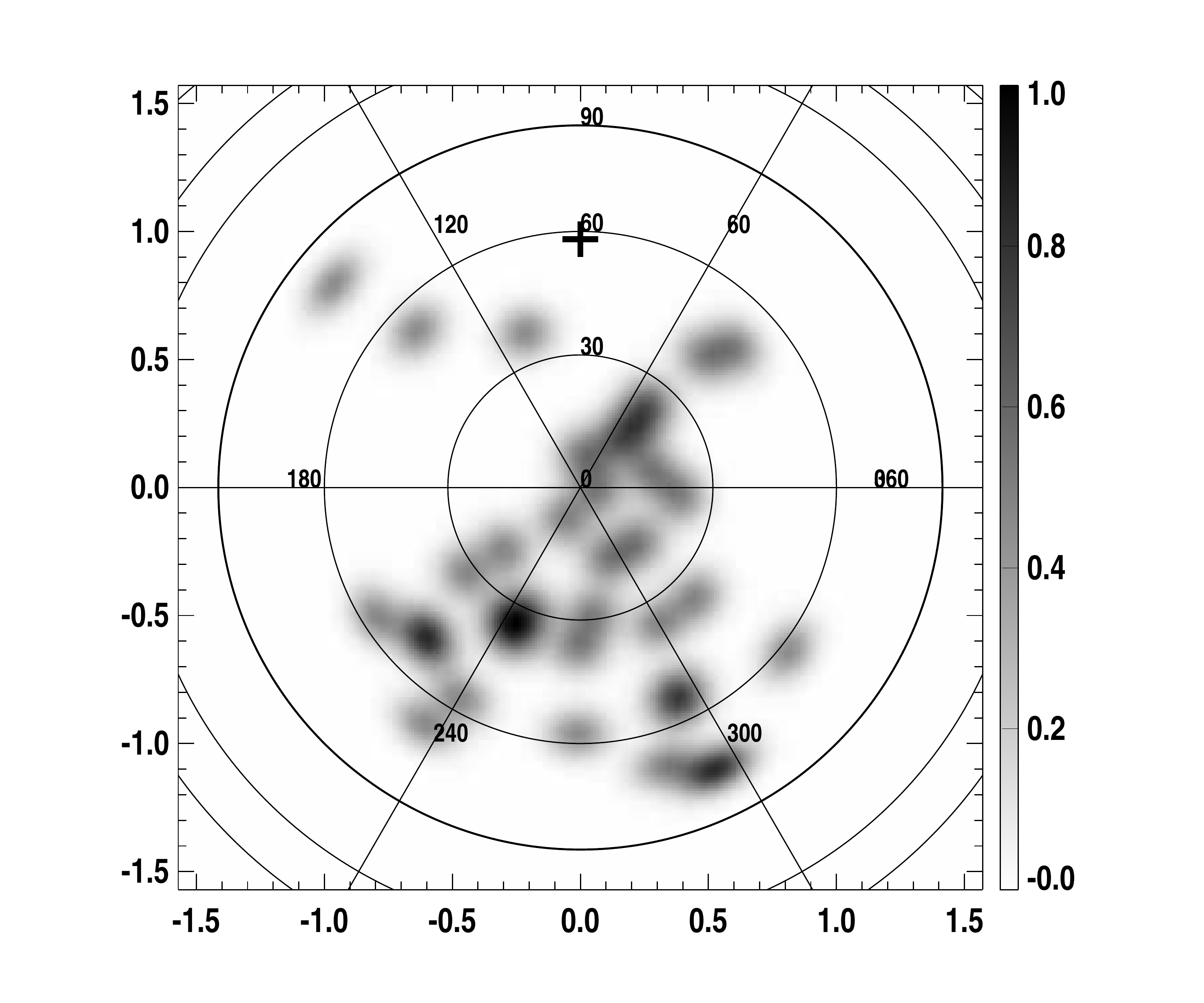}
	\caption{Sky map of 36 radio events registered with the autonomous system in coincidence with the SD Array in local spherical coordinates. The zenith is at the center and the azimuths are oriented as follows: $0^\circ$ east, $90^\circ$ north. The geomagnetic field direction in Malarg\"ue is indicated by the cross.}
	\label{skyplot}
	\end{figure}

\noindent	
The sky distribution of radio events observed with the setup near the CLF and in coincidence with the SD Array is presented in Fig. \ref{skyplot}. As expected we observe that the efficiency for the observation of cosmic rays with the SD Array does not depend on the azimuth angle. The sky plot of the observed radio events, however, is highly asymmetric with a large excess of events arriving from the south: this represents more than 70\% of all events. This observed asymmetry provides further support for the geomagnetic origin of radio emission by air showers and has been observed before in the northern hemisphere \cite{Ardouin2009a}. The Earth's magnetic field vector at the site of the Pierre Auger Observatory makes an angle of about $60^\circ$ with respect to the zenith and its azimuthal angle is $90^\circ$ (i.e. north). This is indicated as the cross in Fig. \ref{skyplot}.  In the geomagnetic model, electric pulses will be strongest if the shower axis is perpendicular to the magnetic field vector. And for these observed events the trigger for radio detection was a simple pulse-height threshold on the radio signal, which explains why we detected more events coming from the south than from the north. \\    
	
\noindent			
Studies with solitary systems have been performed. These systems are powered by solar energy, have a wireless connection to a central DAQ system, and operate in self-trigger and/or externally triggered modes. Special attention has been paid to reduce self-induced noise caused by DC-DC converters and by digital electronics. With this system self-triggered events have been obtained. However, for these events there were no data from more than two radio stations and thus the arrival direction could not be compared with that from SD. Nevertheless this is the first time ever that such an independent detection has been achieved using Radio-Detection Stations \cite{Revenu2008a}.\\

\begin{figure}[h!t]
		\centering
		\includegraphics[width=0.4\textwidth]{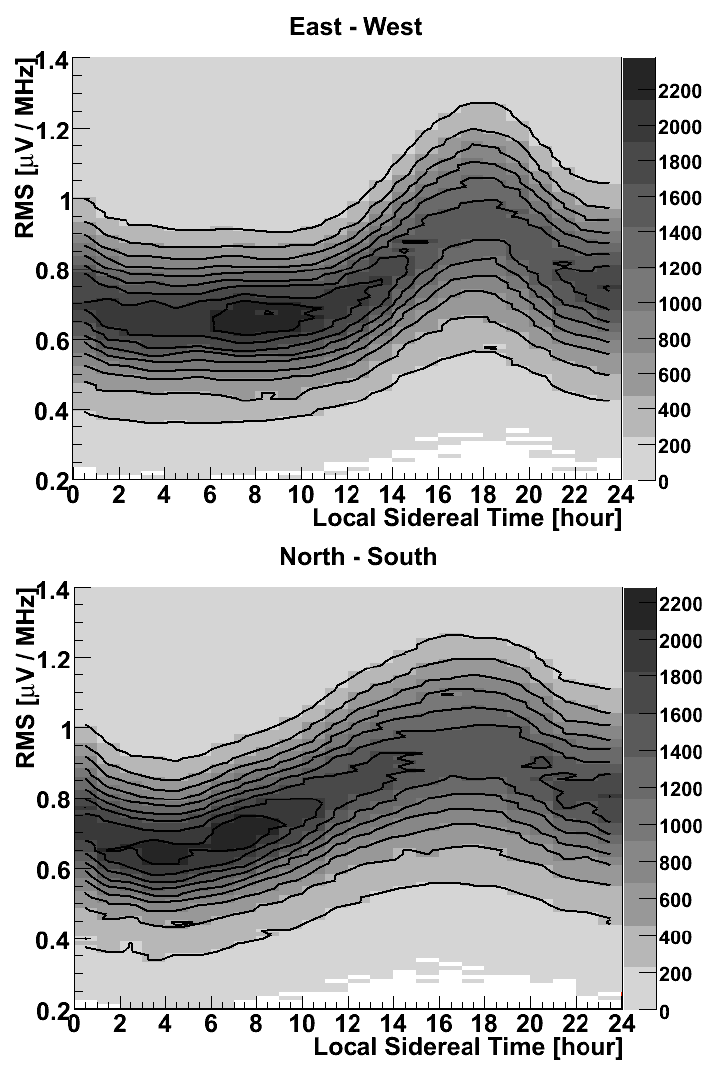}
		\caption{RMS value of the noise level measured with one of the Radio-Detection Stations as a function of the LST filtered for the frequency band between 50 and 55 MHz. The plotted RMS values have been corrected for cable losses and amplification between the output of the logarithmic periodic dipole antenna and the receiver unit, but not for the antenna gain. The grey scale indicates number of counts per hour and per 0.012 $\mu$V~MHz$^{-1}$. The data were collected between May 2007 and April 2008.}
		\label{noise}
	\end{figure}

\noindent
Noise levels have been studied in detail. For this analysis we have defined the root mean square (RMS) value for 750 consecutive samples of each time trace with have a sampling rate of 400 MS~s$^{-1}$. Using a digital filter the analyzed signals were limited to the frequency band between 50 and 55~MHz. The distribution of these RMS values, measured for a period of almost one year, are plotted in Fig. \ref{noise}. This figure displays for two polarizations the distribution of all the calculated RMS values as a function of the local sidereal time (LST) for one of the Radio-Detection Stations. For both directions of the polarization, we recognize the same trend: an increase in the RMS value around 18:00 LST (see also  Ref. \cite{Coppens2008a}). This time coincides with the passage over the Radio-Detection Station of an area near the galactic center with an increased brightness temperature in the radio frequency band \cite{radiosources}. These studies enable us to monitor the overall detection threshold and the gain pattern of antennas used. \\

\noindent
The signal development of radio pulses induced by UHE CRs in the atmosphere has been modeled using extensive Monte-Carlo simulations, tracking electron- and positron density distributions, and by using macroscopic calculations \cite{Huege2009a}. In addition, a modular detector simulation package has been written, where various important components of the signal development as it passes through the electronics chain can be simulated and compared with data \cite{Fliescher2008a}. A conceptual design has been made for the offline analysis and visualization codes \cite{Rautenberg2008a}. This package will be merged with the standard off\-line analysis which is being used for the analysis of data obtained with the other detector systems used in the Observatory.\\

		\begin{figure*}[h]
		\centering
		\includegraphics[width=0.875\textwidth, viewport = 50 75 680 325, clip]{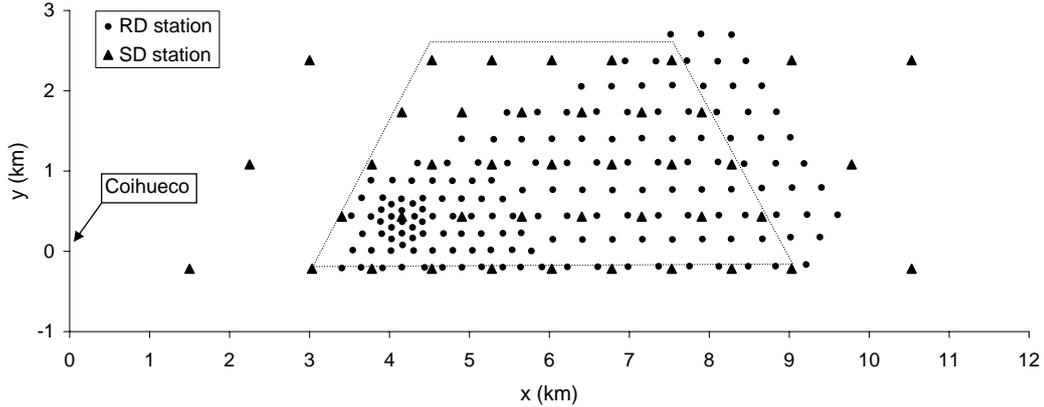}
		\caption{Areal view of AERA at the western part of the Pierre Auger Observatory. Radio-Detection Stations are denoted as dots, detectors of the SD Array as triangles. All coordinates are relative to the Coihueco fluorescence building; see also Fig. \ref{layoutAS}.}
		\label{layout}
		\end{figure*}

\section{AERA: the 20 km$^{2}$ array}
		
The baseline parameters for AERA are about 150 Radio-Detection Stations distributed over an area of approximately 20 km$^2$. Figure \ref{layout} shows an areal view of the site. AERA will have a core of 24 stations deployed on a triangular grid with a baseline of 150~m. This core is about 4~km east of the Coihueco fluorescence telescope; it provides an excellent overlap for events which will be observed with both detection systems. Around this core, there will be 60 stations on a triangular grid with a pitch size of 250~m. Finally, the outer region of AERA will have 72 stations with a mutual distance of 375~m.  Each Radio-Detection Station will operate on solar power and has its local data-acquisition. Like the SD of the Observatory, event definition will be based on timing information sent through wireless communication by the Radio-Detection Stations to a central data-acquisition system, located near the center of AERA. The design of the antennas and of the electronics will be optimized to have a high sensitivity in the frequency band between 30 and 80~MHz. Every Radio-Detection Station will have a ring buffer which can contain streaming data (4 channels, 200 MS s$^{-1}$, 12 bits per sample) for a period of 3 s. \\

	\begin{figure}[h!t]
		\centering
		\includegraphics[width=0.4\textwidth, viewport = 0 0 230 180, clip]{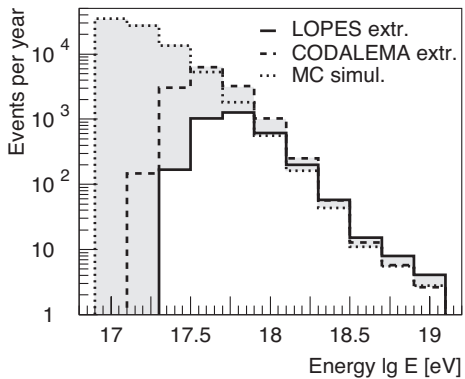}
		\caption{Expected number of events per year as function of the shower energy according to extrapolations of LOPES and CODALEMA measurements as well as Monte-Carlo simulations, based on the REAS2 code and with zenith angle $\Theta<60^\circ$; see text for details.}
		\label{eff}
	\end{figure}

\noindent
The expected performance of the array has been calculated applying different methods. Measurements of the lateral distribution of the radio signal as obtained by the LOPES \cite{Horneffer2007a} and CODALEMA \cite{codalema_unpublished} experiments have been taken into account as well as shower simulations with the REAS2 code \cite{Huege2007a}. To estimate the expected event rates, the effective area is multiplied with the cosmic-ray flux. The present estimates are based on the spectrum measured by the Pierre Auger Collaboration \cite{Abraham2008a}. The flux below $10^{18.45}$~eV has been extrapolated, assuming a spectral index of $-3.3$, which is typical for the energy region between the second knee and the ankle, $10^{17.6}$ - $10^{18.6}$~eV. The expected number of events per year is depicted in Fig. \ref{eff} for the three different approaches. The main difference between the approaches is in the threshold region, where the largest uncertainty comes from the extrapolation of the flux spectrum to lower energies. Taking a conservative approach and using the CODALEMA and LOPES extrapolations, the threshold for the radio array is around $\approx 10^{17.2}$~eV.\\

\noindent
Summarizing, AERA will be the first large radio-detection array for the observation of UHE CRs. As it will be co-located with the other detector systems of Auger it will provide additional and complementary information on air showers which can be used, e.g., for the determination of the composition of UHE CRs.

\end{document}